\newcommand{\al}{\mbox{$^{26}$\hspace{-0.2em}Al}}

\newcommand{\Msol}{\mbox{$M_{\sun}$}}

\def\MeV{\mbox{Me\hspace{-0.1em}V}}

\def\sun{\hbox{$\odot$}}
\def\deg{^\circ}

\documentstyle[epsfig]{aipproc}
\begin{document}

\title{\al\ sources in the Galaxy as seen in the 1.809 \MeV\ gamma-ray line}
\author{J.~Kn\"odlseder$^{\ast}$}
\address{$^{\ast}$Centre d'Etude Spatiale des Rayonnements, B.P.~4346, 31028 Toulouse 
         Cedex 4, France}
\maketitle

\begin{abstract}
Gamma-ray line observations provide a versatile tool for studies of 
nucleosynthesis processes and supernova physics.
In particular, the observation of radioactive species in the 
interstellar medium probes recent nucleosynthesis activity on various 
time-scales for different kinds of sources.
Considerable progress in gamma-ray instrumentation during the last 
decades has led to the discovery of several cosmic gamma-ray lines.
The best studied of these lines is today the 1.809 \MeV\ line attributed to 
the decay of radioactive \al\ within the interstellar medium.
In this review, recent observational results are presented and their 
astrophysical implications are discussed.
\end{abstract}

\section*{Introduction}

Convincing proof of ongoing nucleosynthesis in the Galaxy comes from 
the observation of the 1.809 \MeV\ gamma-ray line, emitted during the 
radioactive decay of \al.
With a lifetime of $10^6$ yr, much shorter than the timescale of 
galactic evolution, \al\ must have been freshly synthesised to be 
observable in the interstellar medium today.
In principle, \al\ could be produced in appreciable amounts by a 
variety of sources, such as 
massive mass losing stars (mainly during the Wolf-Rayet phase), 
Asymptotic Giant Branch stars (AGBs),
novae (mainly of ONe subtype),
and core collapse supernovae.
In addition to stellar production, \al\ can also be produced by 
spallation reactions of high-energy cosmic rays, although at 
substantially lower rates (see \cite{prantzos96} for a recent review).

Considerable uncertainties that are involved in the modelling of 
nucleosynthesis processes, mainly due to the poorly known physics of 
stellar convection, do not allow for a theoretical determination of 
the dominant galactic \al\ sources.
In view of this difficulty, is has been suggested that help could be 
expected from a combination of improved spatial source distribution 
models and observations with good angular resolution \cite{leising85}.
However, until recently, the angular resolution and sensitivity of 
gamma-ray detectors were too poor to perform such a mapping; hence the 
origin of galactic \al\ remained veiled.

The situation changed dramatically after the launch of the Compton 
Gamma-Ray Observatory ({\em CGRO}), in April 1991.
The COMPTEL telescope aboard {\em CGRO} performed the first mapping of the 
Galaxy in the light of 1.809 \MeV\ photons \cite{diehl95}.
I will review these observations in this paper, and summarise what 
has been learned about the origin of \al.

\section*{Observations}

The COMPTEL telescope allows the study of 1.8 \MeV\ gamma-ray line 
emission with an energy resolution of $\sim8\%$ (FWHM) and an 
angular resolution of $3.8\deg$ (FWHM) within a wide field of view of 
about 1 steradian \cite{schonfelder93}.
Combination of observation periods of typically 1-2 weeks allows to 
extend the field of view to the entire sky, enabling an all-sky 
analysis of cosmic 1.8 \MeV\ line emission.
Using data from the first five years of the {\em CGRO} mission led to 
a point source sensitivity of $1\,10^{-5}$ ph cm$^{-2}$ s$^{-1}$ 
($3\sigma$), a performance that exceeds by far the limits of precedent 
telescopes.
Hence for the first time, galactic 1.8 \MeV\ gamma-ray line emission 
could be studied in great detail.

\begin{figure}[t!]
\epsfig{file=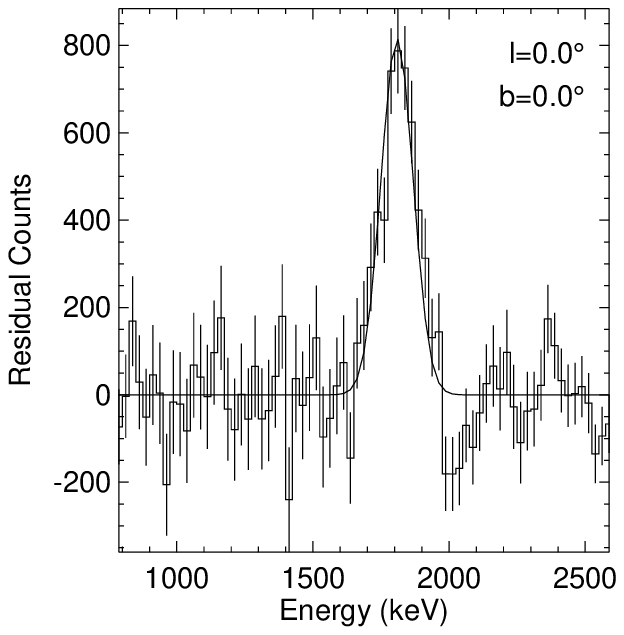,width=4.8cm}
\hfill
\epsfig{file=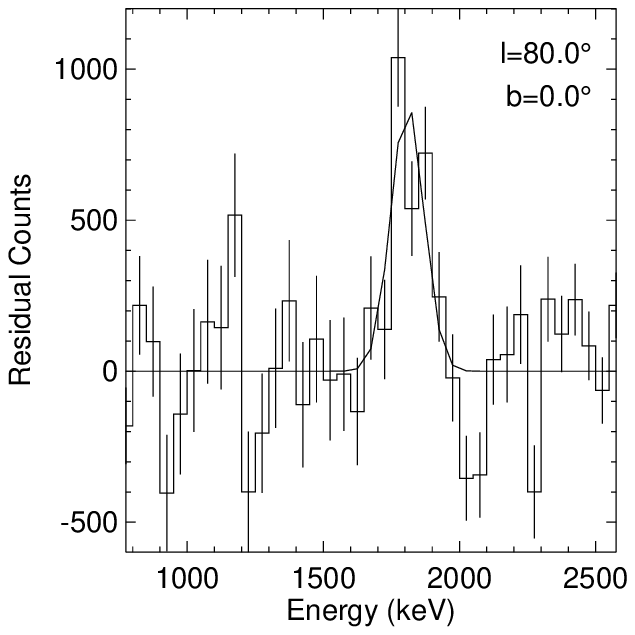,width=4.8cm}
\hfill
\epsfig{file=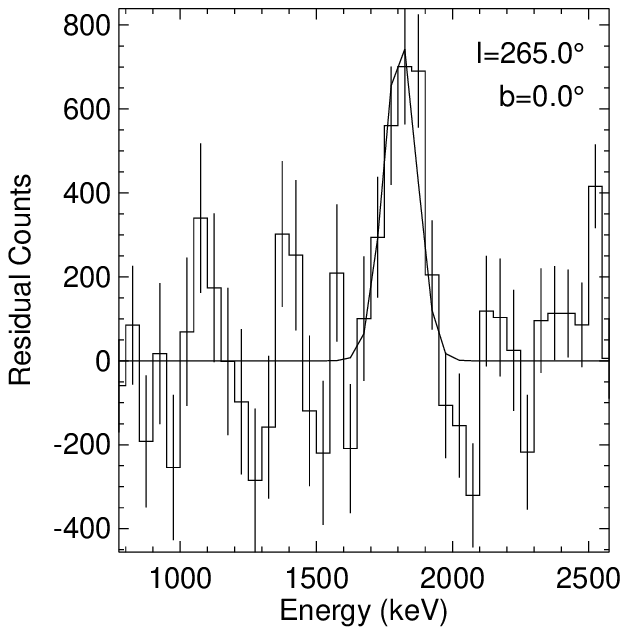,width=4.8cm}
\caption{Background subtracted COMPTEL count spectra of the galactic centre 
(left), the Cygnus region (middle), and the Vela region (right). 
1.809 \MeV\ gamma-ray line signatures are clearly visible in all 
spectra.}
\label{fig:lines}
\end{figure}

For illustration, background subtracted COMPTEL count spectra for 
three regions along the galactic plane are shown in Fig.~\ref{fig:lines}
\cite{knoedl97}.
The first panel shows the well known 1.8 \MeV\ line from the galactic centre 
region which has been detected by numerous instruments 
\cite{prantzos96}, but never at such high significance.
1.8 \MeV\ line emission from Cygnus and Vela, as seen in the next two 
panels, has never been reported before.
Both regions are roughly $90\deg$ away from the galactic centre, 
illustrating that \al\ production is not only concentrated on the 
galactic centre.
In fact, from the relative intensities with respect to the galactic 
centre emission it becomes clear that the galactic \al\ distribution 
obeys a rather flat profile along the galactic plane, inconsistent 
with proposed distributions for novae that are highly peaked towards 
the galactic centre.

\begin{figure}[t!]
\epsfxsize=14.8cm \epsfclipon
\epsfbox{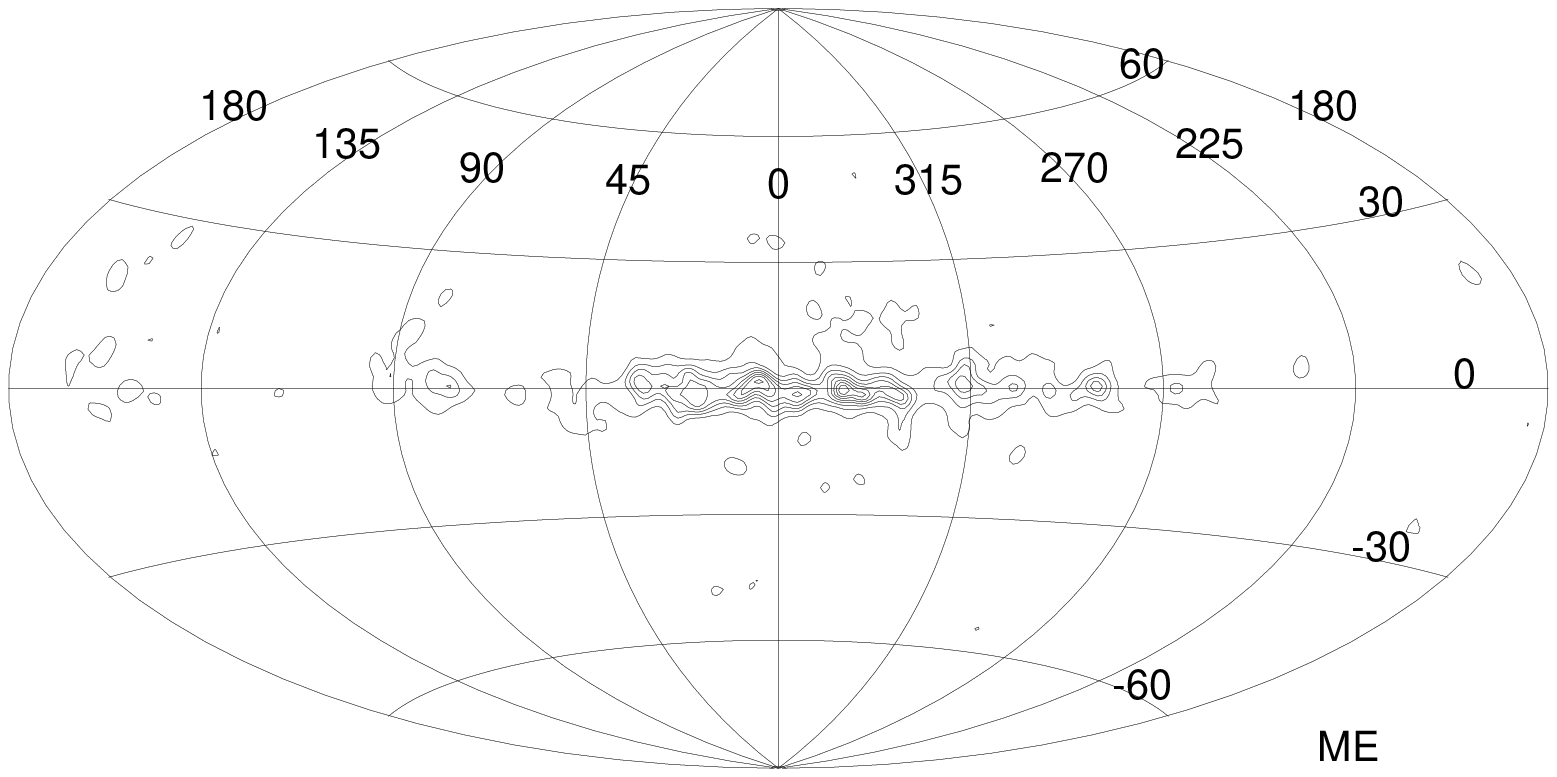}
\hfill
\epsfxsize=14.8cm \epsfclipon
\epsfbox{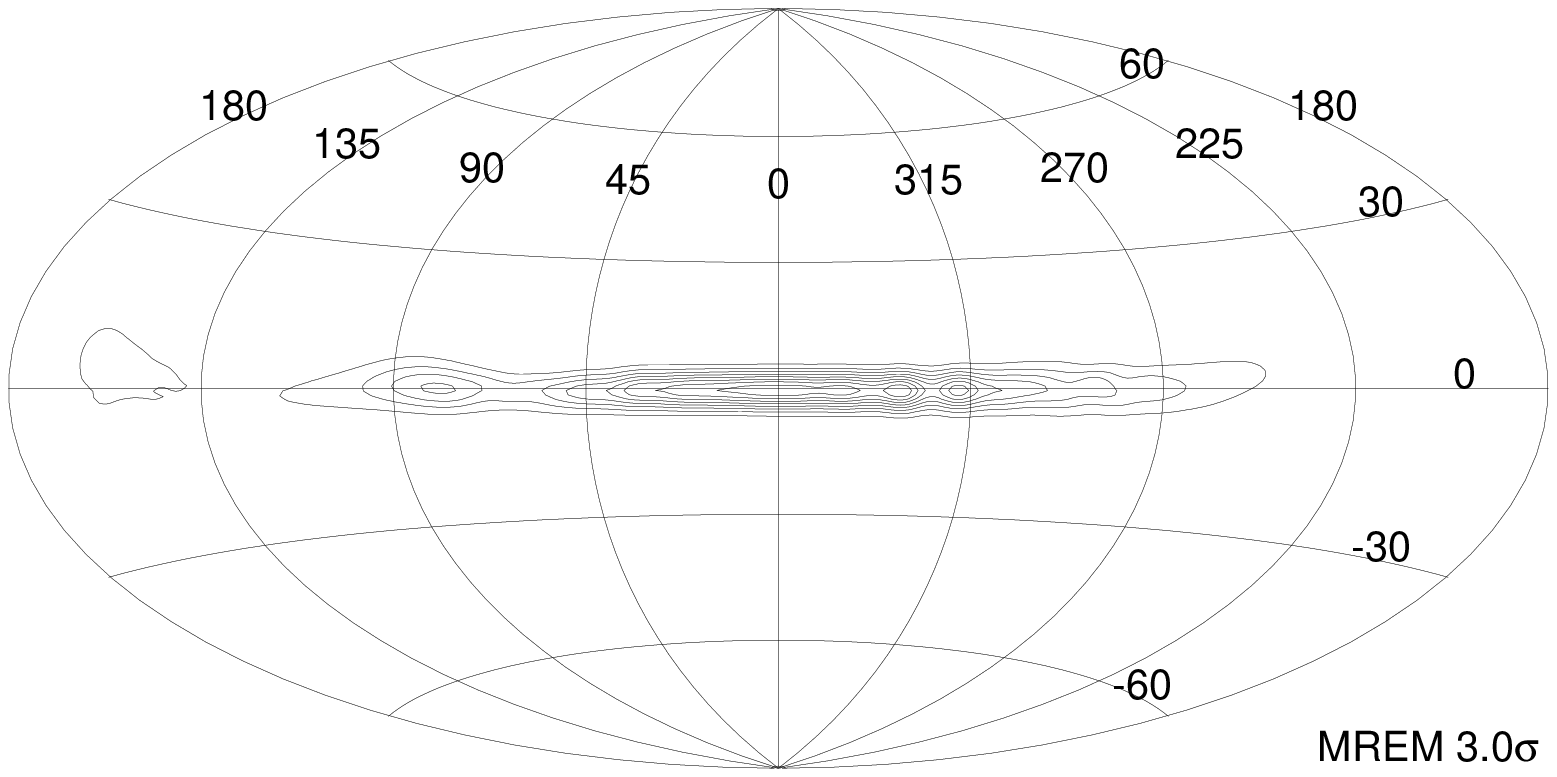}
\caption{COMPTEL 1.8 \MeV\ all-sky maps derived using a maximum 
entropy algorithm (top) and the multi-resolution algorithm MREM 
(bottom).}
\label{fig:al26}
\end{figure}

COMPTEL images of 1.8 \MeV\ gamma-ray line emission are obtained by 
deconvolving the data in a narrow energy band centred on the line 
energy.
This procedure relies on accurate modelling of the instrumental 
background which is obtained by interpolating observations at adjacent 
energies, thus removing possible contributions from continuum sources 
\cite{knoedl96a}.
A maximum entropy (ME) all-sky image of 1.8 \MeV\ line emission is shown in 
Fig.~\ref{fig:al26} in Aitoff-projection \cite{oberlack96}.
The image reveals lumpy emission structures along the galactic plane 
with `hot spots' eventually separated by emission-free `gaps'.
Some degree of lumpiness in the 1.809 \MeV\ emission is expected if massive 
stars are at the origin of galactic \al, since they mainly form in stellar 
associations which are probably aligned along the spiral arm structure 
of the Galaxy.
Indeed, the comparison of the data to detailed models of galactic 
spiral structure supports such a correlation \cite{knoedl96b}. 
However, the amount of emission lumpiness seen in the ME image even 
exceeds expectations for massive star associations \cite{lentz99}, 
questioning the origin of the image lumpiness.

It has gradually become clear that at least part of the image 
lumpiness is an artifact of the image reconstruction process 
\cite{knoedl96a}.
It results from the weak constraints that are imposed on individual 
image pixels by our data, leading to a propagation of statistical 
noise into the deconvolved skymaps.
We therefore developed a new algorithm, called MREM, that we 
specifically designed for the reconstruction of diffuse, low-level 
gamma-ray emission \cite{knoedl99b}.
Application of the MREM algorithm to COMPTEL data leads to an alternative 
image of 1.809 \MeV\ gamma-ray line emission that reveals much less 
image lumpiness (cf.~Fig.~\ref{fig:al26}).
We believe that most emission structures that are seen in the MREM image 
are indeed real.
However, we cannot exclude that weak emission structures that are close 
to the sensitivity limit of COMPTEL have been smoothed out by the 
MREM reconstruction process.

Despite the different amounts of image lumpiness in the two
reconstructions, both skymaps have features in common.
The maps reveal an intense, asymmetric ridge of diffuse 1.809 \MeV\ emission 
along the galactic plane with a prominent localised emission 
enhancement in the Cygnus region.
The longitude profile of the emission is rather flat and does not show 
any pronounced emission enhancement towards the galactic centre.
This clearly illustrates that \al\ nucleosynthesis is not localised 
to specific areas of our Galaxy, such as the galactic centre or the 
local interstellar medium \cite{hillebrandt87,morfill85}.
It is a galaxywide phenomenon which reflects the recent nucleosynthesis activity 
throughout the Milky Way.

The emission enhancement in Cygnus is of particular interest.
By searching for possible counterparts of the emission, it turned out 
that the observations can be explained by the combined nucleosynthesis 
activity of Wolf-Rayet stars and core collapse supernova in a group of 
OB associations and young open clusters situated at 1-2 kpc from the 
Sun \cite{delRio96}.
Hence, the Cygnus feature is a fingerprint of recent nucleosynthesis 
activity in the local interstellar medium, probably related to the 
local spiral arm structure.
In this picture, the 1.8 \MeV\ emission seen towards Vela could represent
the continuation of the local spiral arm at negative longitudes after 
passing near the Sun \cite{knoedl97b}.

In our continuous effort in finding tracers of recent nucleosynthesis 
activity \cite{diehl95,diehl97} we recently discovered two 
intensity distributions that are closely correlated to 1.809 \MeV\ 
emission: 
galactic free-free emission, observed by the DMR telescope aboard {\em COBE} 
in the microwave domain, and the 158 $\mu$m fine-structure line of C$^+$, 
observed by FIRAS (also aboard {\em COBE}) \cite{knoedl99a,knoedl99c}.
Both distributions are tracers of the ionised interstellar medium, 
hence they directly reflect the galactic population of very massive 
(M$_{i} > 20$ \Msol) stars \cite{knoedl99}.
The observed correlations suggest that \al\ is primarily produced by the 
same population.
Consequently, due to the short lifetime of massive stars, \al\ 
becomes an excellent tracer of recent galactic star formation.

From the correlations I inferred an equivalent O7V star \al\ yield of 
$Y_{26}^{O7V}=(1.0\pm0.3)\,10^{-4}$ \Msol, expressing the amount 
of \al\ produced by a typical massive, ionising star.
Remarkably, this value is in good agreement with yield estimates for 
massive Wolf-Rayet stars \cite{langer95,meynet97}.
Using an estimate of the galactic Lyc luminosity of $Q=3.5\,10^{53}$ 
ph s$^{-1}$ \cite{bennett94}, the equivalent O7V star \al\ yield converts 
into a galactic \al\ mass of $3.1\pm0.9$ \Msol.
This value is to be compared to nucleosynthesis predictions of 
$2.2\pm0.4$ \Msol, where $\sim60\%$ of the galactic \al\ is predicted 
to arise from Wolf-Rayet stars while $\sim40\%$ should come from core collapse 
supernovae \cite{knoedl99}.

\begin{figure}[t!]
\centerline{
\hfill
\epsfxsize=14.8cm \epsfclipon
\epsfbox{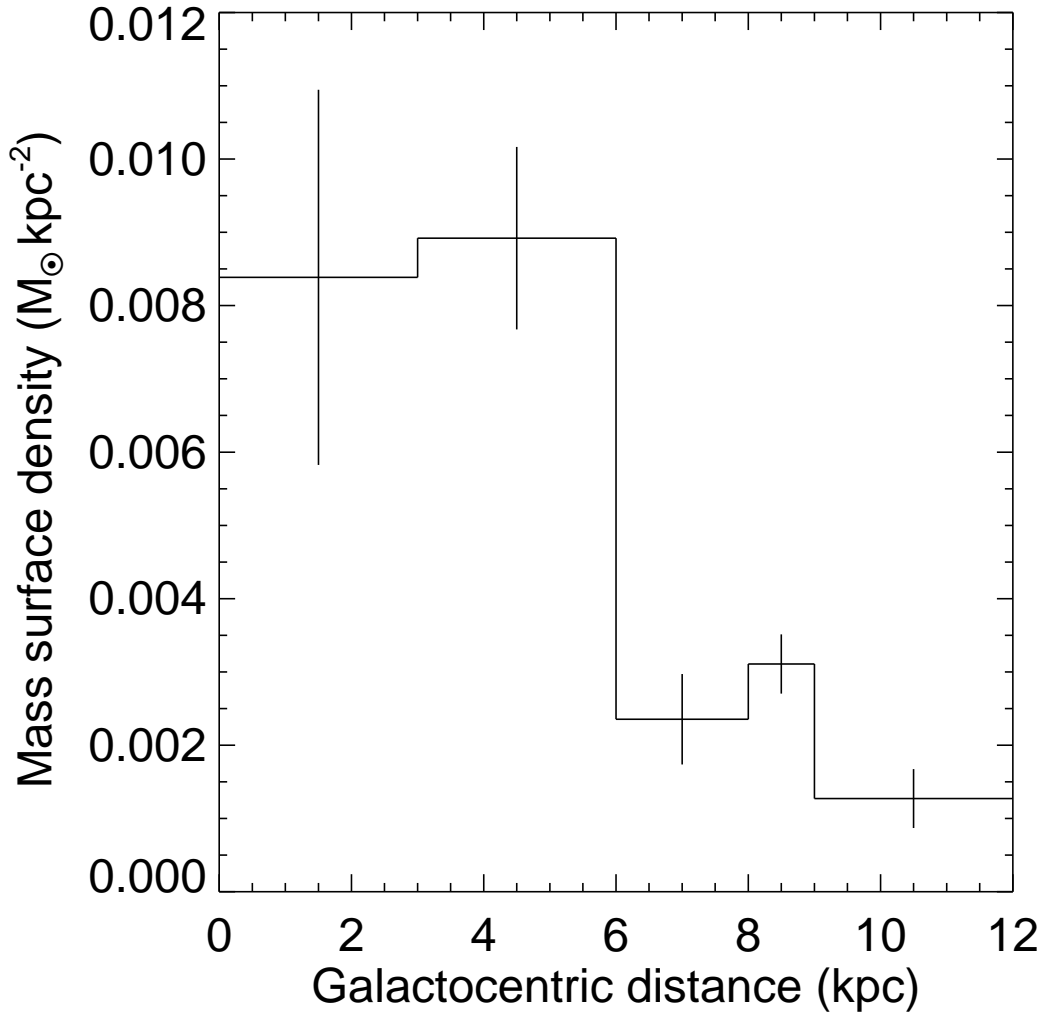}
\hfill}
\caption{\al\ mass surface density as derived from COMPTEL 1.8 \MeV\ 
  gamma-ray line data.}
\label{fig:radprofile}
\end{figure}

We can now use \al\ to learn more about massive star formation in the 
Galaxy by observations of the 1.809 \MeV\ gamma-ray line.
For example, using a maximum likelihood procedure, we can derive the radial 
\al\ mass density distribution in the Galaxy, which in turn reflects 
the radial dependence of the star formation rate.
The resulting profile is shown in Fig.~\ref{fig:radprofile} 
\cite{knoedl97}.
The data illustrate that the bulk of galactic star formation occurs 
at distances of less than 6 kpc from the galactic centre.
Star formation is also present within the central 3 kpc of the Galaxy, 
although at a poorly determined rate.
There are indications for enhanced star formation between $3-6$ kpc, 
coinciding with the molecular ring structure as seen in CO data.
Enhanced star formation is also seen in the solar neighbourhood 
($8-9$ kpc) which probably corresponds to the local spiral arm 
structure.
However, the radial \al\ profile is probably not directly 
proportional to the radial star formation profile since \al\
nucleosynthesis may depend on metallicity.
It will be important to determine this metallicity dependence in order 
to extract the true star formation profile from gamma-ray line data.
Valuable information about the metallicity dependence will come from a 
precise comparison of the 1.809 \MeV\ longitude profile to the profile of 
free-free emission.
Additionally, observations of gamma-ray lines from $^{60}$Fe, an 
isotope that is only believed to be produced during supernova 
explosions, can help to distinguish between hydrostatically and 
explosively produced \al, and therefore help to disentangle 
the metallicity dependencies for the different candidate sources.

\section*{Where do we stand? Where do we go?}

COMPTEL observations of the 1.809 \MeV\ gamma-ray line mark the 
beginning of a new epoch in gamma-ray line astronomy.
For the first time, the entire Galaxy has been imaged in a gamma-ray 
line, supplying us with a map of recent nucleosynthesis activity.
A connection between nucleosynthesis and massive star formation has 
been established, linking gamma-ray line astronomy to fields like 
stellar evolution, galactic evolution, galactic structure, and the 
physics of the interstellar medium.

COMPTEL observations suggest that massive stars are the primary source 
of galactic \al, although a small contribution (10-20\%) from novae 
or AGB stars cannot be excluded at this point.
The expectations are now focused on SPI, the spectrometer aboard 
the {\em INTEGRAL} observatory, scheduled for launch in 2001.
SPI not only brings a further improvement in sensitivity and angular 
resolution, it also combines imaging with high-resolution 
spectroscopy, a novelty in gamma-ray line astronomy
(see {\tt 
http://astro.estec.esa.nl/SA-general/Projects/Integral/
integral.html}).
During the core program, which will cover $25-35\%$ of the observation 
time, INTEGRAL will perform a deep survey of the central radian and a 
weekly scan of the galactic plane.
During these surveys, significant exposure will be accumulated along the 
galactic plane, allowing for a detailed mapping of galactic 1.809 \MeV\ 
emission.
Additionally to the spatial mapping, line shapes will be measured to 
high precision, enabling to detect line broadenings or shifts of a few 
100 km s$^{-1}$.
The line shape analysis is of particular importance in light of 
a possible 1.809 \MeV\ line broadening, suggested by {\em GRIS} 
balloon observations of the galactic centre region \cite{naya96}.

The remaining observation time is open to the scientific community, 
allowing for detailed studies of localised 1.809 \MeV\ emission 
features.
A prime candidate for such studies is, of course, the Cygnus region, 
where SPI observations could reveal substructures related to different 
massive star populations.
Comparison of these observations with other tracers of massive star 
formation, such as galactic free-free emission, and their 
interpretation using multi-wavelength evolutionary synthesis models, 
will provide a powerful analysis tool for studies of nucleosynthesis 
processes in this region \cite{knoedl00}.
Comparable in interest are probably the Vela and the Carina regions, 
where hints for localised emission features have been found in COMPTEL 
data.

Thus, the story about galactic 1.809 \MeV\ is by far not over -- it 
just began.
It is not unrealistic that 1.809 \MeV\ line diagnostics will become a 
standard tool in astrophysics, complementing information obtained at 
other wavelengths.
Stay tuned \ldots


\end{document}